\documentclass[twocolumn]{emulateapj}

\usepackage{epsfig}
\usepackage{amssymb}

\newcommand{\msun}{\mbox{$M_\odot$}}
\newcommand{\rsun}{\mbox{$R_\odot$}}
\def\be{\begin{eqnarray}}
\def\ee{\end{eqnarray}}
\def\lsim{\mathrel{\rlap{\lower3pt\hbox{\hskip1pt$\sim$}}
     \raise1pt\hbox{$<$}}} 
\def\gsim{\mathrel{\rlap{\lower3pt\hbox{\hskip1pt$\sim$}}
     \raise1pt\hbox{$>$}}} 

\begin{document}

\shorttitle{Kerr Parameters of Black Holes}
\shortauthors{Moreno M\'endez, et al.}

\title{Kerr Parameters for Stellar Mass Black Holes and Their Consequences for GRBs and Hypernovae}

\author{Enrique Moreno M\'endez, Gerald E. Brown}
\affil{Department of Physics and Astronomy,
               State University of New York, Stony Brook, NY 11794, USA.}

\email{EMM: moreno@grad.physics.sunysb.edu}

\email{GEB: gbrown@insti.physics.sunysb.edu}

\author{Chang-Hwan Lee}
\affil{Department of Physics, Pusan National University,
              Busan 609-735, Korea.}
\email{CHL: clee@pusan.ac.kr}

\and
\author{Frederick M. Walter}
\affil{Department of Physics and Astronomy,
               State University of New York, Stony Brook, NY 11794, USA.}


\begin{abstract}

Recent measurements of the Kerr parameters $a_\star$ for two black-hole binaries in our Galaxy \citep{Sha06},
GRO J1655$-$40 and 4U 1543$-$47 of $a_\star=0.65-0.75$ and $a_\star=0.75-0.85$, respectively, fitted well the predictions of \citet{Lee02}, of $a_\star\cong 0.8$.  In this report we also note that \citet{Lee02} predicted $a_\star>0.5$ for 80\% of the Soft X-ray Transient Sources. The maximum available energy in the Blandford-Znajek formalism for $a_\star > 0.5$ gives $E > 3\times10^{53}$ergs, orders of magnitude larger than the energy needed for the GRB and hypernova explosion.
We interpret the Soft X-ray Transients to be relics of GRBs and Hypernovae, but most of them were subluminous ones which could use only a small part of the available rotational energy.
%
%
\end{abstract}



\section{Introduction}\label{sec-Intro}

Recent estimates of the Kerr parameters $a_\star$ for two Soft X-ray Transients (SXTs) \citep{Sha06}, GRO J1655$-$40 (Nova Sco) and 4U 1543$-$47 (Il Lupi), facilitate a test of stellar evolution, in that the spins of the black holes in these binaries should be produced in common envelope evolution which begins with the evolving massive giant and companion donor, and ends up in helium-star--donor binary, the hydrogen envelope of the massive star having been stripped off and the helium having been burned.

\begin{figure}
\centerline{\epsfig{file=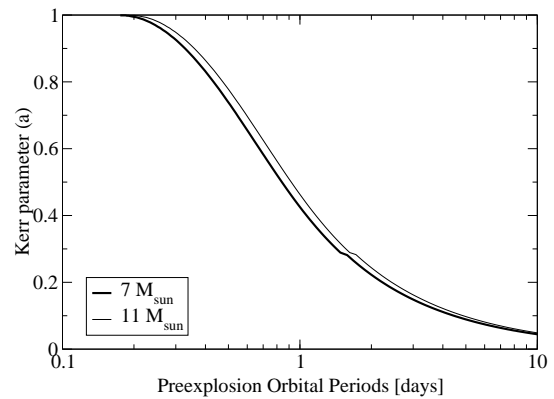,height=2.5in}} \caption{The Kerr parameter of the black hole resulting from the collapse of a helium star synchronous with the orbit, as a function of orbital period (LBW). Thick (thin) solid line corresponds to initial $7\msun$ ($11\msun$) He star. Note that the result depends very little on the mass of the helium star.
\label{FIG12} }
\end{figure}

\citet{Lee02} (hereafter denoted as LBW) assumed common envelope evolution to begin only after He core burning has been completed; i.e., Case C mass transfer \citep{Bro01b}. Otherwise the He envelope, if laid bare, would blow away to such an extent that the remaining core would not be sufficiently massive to evolve into a black hole \citep{Bro01a}.
%
%
The black-hole-progenitor star, in which the helium core burning has been completed, is tidally locked with the donor (secondary star) so the spin period of the helium star is equal to the orbital period of the binary. In this tidal locking, LBW assumed uniform rotation of He star by assuming that the inner and outer parts of He are strongly connected due to the presence of a strong magnetic field.
The C-O core of the helium star drops into a rapidly spinning black hole due to angular momentum conservation.
In this process, the spin of the black hole depends chiefly on the mass of the donor because the orbital period chiefly depends on the donor mass as we explain later (in section~\ref{sec-LMD2HAM}).
%
%
LBW calculated this Kerr parameter ($a_\star$) as a function of binary orbital period. The results are given in their Fig.~12 which we reproduce as Fig.~\ref{FIG12}.
The agreement of the natal Kerr parameters with the \citet{Sha06} measured ones means that only a small amount of angular momentum energy could have been lost after the formation of the black hole.  The good agreement in this comparison supports the assumption of Case C mass transfer and the tidal locking at the donor-He star stage assumed in the LBW calculations.

In the LBW calculation, the $a_\star$ for Nova Sco was slightly greater than for Il Lupi.  Combining this with observation, $0.75<a_\star<0.85$ would be our best estimate for both binaries.
In this report we also note that LBW predicted $a_\star>0.6$ for 7 Soft X-ray Transient Sources with main sequence companions and $a_\star\cong 0.5$ for XTE J1550$-$564 and GS 2023$+$338 (V404Cygni) with evolved companions.

The maximum available energy in the Blandford-Znajek formalism for $a_\star > 0.5$ gives $E > 3\times10^{53}$ergs, orders of magnitude larger than observed in the GRB and hypernova explosion.
Based on this observation, we interpret the Soft X-ray Transients to be relics of GRBs and Hypernovae.
It should be noted that the way in which the hypernova explodes can be similar to the Woosley Collapsar model. The main advantage in our scenario is that the H envelope in our binary is removed by the donor and the rotational energy is naturally produced in the common envelope evolution.  The necessity for Case C mass transfer, given
Galactic metallicity, and the measured system velocity lock us into the Kerr-parameter values we find.




In section~\ref{CaseC} we elaborate on the determination of the Kerr parameters of soft X-ray transient black-hole binaries discussed above, which could be read off the figures of LBW by a discussion of tidal locking, and the connection of tidal locking to Case C mass transfer. We also discuss the energetics for GRBs and Hypernovae based on the black-hole spin.
%
%
In section~\ref{sec-LMD2HAM} we show that the angular-momentum energy of the black-hole binary is determined mainly by the mass of the donor.  We discuss 12 Galactic transient sources with angular-momentum energies $\ge 10^{53}$ergs, so that all of these are relics of GRBs and Hypernovae.  The energies of the GRB and Hypernova explosion powered by these, as we shall develop, should be subtracted from the natal rotational energies, to give the explosion energy.


\section{Case C Mass Transfer and Tidal Locking}\label{CaseC}

Case C mass transfer implies that the mass transfer takes place late, after the He in the giant progenitor of the black hole has been burned.
The proof of our scenario was given in the measured Kerr parameters for Nova Sco and Il Lupi \citep{Sha06} which agreed with the predictions of LBW.  Again, aside from the fact that we use the tidal coupling of the donor to spin of the black hole progenitor, the rest of our scenario, especially the collapse, is the same as in the Woosley Collapsar model. 

%
%

The GRB and Hypernova explosions are all of type $I_c$, so that He lines do not appear.
%
In the Woosley Collapsar model, He burning is not necessarily finished before explosion, but a) the interacting He may fall into the black hole or, b) the He may not mix with the $^{56}$Ni, so that in either case He lines would not be seen.

We remark here that the explosion on which a record number (119) of astronomers concentrated their attention \citep{You06,Cam06,Pia06,Sod06,Maz06}, SN2006aj was a $I_{bcd}$ explosion; i.e. (see Additional Information on \citet{Maz06})  convective carbon comes to an end just at the ZAMS mass at which black holes begin to form, as explained in \citet{Bro01a} (see fig.1 of that paper) because the entropy loss by neutrino emission during convective carbon burning is shut off by the absence of carbon and the entropy that increases with ZAMS mass must go into adding nucleons to the iron core of the star, before it collapses.  The $I_{bcd}$ nature of the explosion identifies the central engine, therefore, as a black hole.  In the 4 Nature papers focussed on GRB060218/SN2006aj the authors speculated on the central engine being a magnetar, but, in fact, it must have been a black hole and the GRB and Hypernova run by the Blandford-Znajek mechanism.  Otherwise there would have been carbon lines.

In \citet{Bro00} the black hole formation was described by a Blaauw-Boersma explosion, which should be sufficient for calculating the system velocity of the binary because conservation laws are respected.  However, we believe the Woosley Collapsar model to give a more detailed description of the black hole formation and the hypernova explosion.  The \citet{Mac99} description includes magnetohydrodynamics in the form of the Blandford-Znajek mechanism.  However, mass loss in the explosion and conservation laws are those of Blaauw-Boersma.


\subsection{Soft X-ray Transients as relics of GRBs and Hypernovae}\label{sec-NSco}

LBW found that there are two classes of soft X-ray transients, those with main sequence companions\footnote{Although they are called main sequence, the companions are mostly highly evolved K-stars.} (denoted as AML), and others with evolved companions (denoted as Nu). Due to the angular momentum loss, via gravitational wave radiation and magnetic braking, after black-hole formation the orbits of AMLs are shortened. Based on this argument and the current observation, LBW traced back the orbital period at the time of black-hole formation in their Fig.~10.  By the tidal locking, the estimated Kerr parameters for AMLs are $a_\star >0.6$ (about half of them are $a_\star >0.8$). The maximum available energies for these system via the Blandford-Znajek formalism are $E > 3\times 10^{53}$ ergs.
So we believe that those black-hole binaries with main sequence companions are the relics of GRBs and Hypernovae.

The evolution of Nu's after black-hole formation is mainly controlled by the donor which is evolving beyond the main-sequence stage, and the orbit is widened due to the conservative mass transfer from the less massive donor to the black hole. We will discuss the possibilities of Nu's as relics of GRBs and Hypernovae.

We will be brief in reconstructing Nova Sco as a relic of GRB and Hypernova because this was constructed in considerable detail with assumed Kerr parameter of $a_{\star}=0.8$ in  \citet{Bro00}.  With the Smithsonian-Harvard-MIT measurement of $a_{\star}=0.65-0.75$ we don't need to change the Brown et al. discussion by much.  Indeed, the Smithsonian-Harvard-MIT $a_{\star}$ gives its value after powering the GRB and Hypernova explosion, whereas the \citet{Bro00} assumed value was before the explosion, so the two are not significantly different, since the explosion can be powered by the energy from an $\sim5\%$ change in $a_\star$ when $a_\star$ is large.

The hypernova aspect of the explosion in Nova Sco was clear by the accretion of $\alpha$-particle nuclei onto the donor as a result of the hypernova explosion.  \citet{Isr99} found that O, Mg, Si and S have abundances on the F-star donor $6-10$ times solar.  These nuclei were presumably absorbed by the donor, which acts as witness to the explosion.

\begin{figure}
\centerline{\epsfig{file=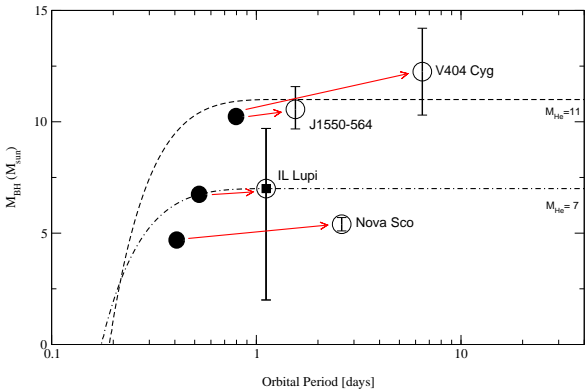,height=2.5in}}
\caption{Reconstructed pre-explosion orbital period vs.\ black hole masses of SXTs with evolved companions. The reconstructed pre-explosion orbital periods and black hole  masses are marked by filled circles, and the current locations of binaries with evolved companions are marked by open circles. The solid lines are ideal polytropic He stars, but both IL Lupi and Nova Scorpii were evolved from $11\msun$ He stars. This figure is obtained from
Fig.~11 of LBW.} \label{FIG11}
\end{figure}

Due to the similarity in the orbital period between Nova Sco and IL Lupi as summarized in Fig.~\ref{FIG11} (LBW),
we argue that IL Lupi is also a relic of GRB and Hypernova.  Although the black hole masses are not known as well in XTE J1550$-$564 and GS 2023$+$338 (V404 Cygni), it can be seen from Fig.~\ref{FIG12} and Fig.~\ref{FIG11} that using their reconstructed preexplosion periods they have a Kerr parameter of $a_{\star}\cong0.5$, possibly somewhat less definite than the prediction of the $a_{\star}$ for Nova Sco.
The latter two binaries have black holes with nearly double the mass as the first two, and, therefore, larger accretion disks.  From our arguments, in Appendix~\ref{app2}, we believe that they may be able to accept more rotational energy, which could be checked by subtracting the measured Kerr parameters from the natal ones.

\citet{Bro00} remarked that for Nova Sco {\it ``After the first second the newly evolved black hole has $\sim10^{53}$erg of rotational energy available to power these.  The time scale for delivery of this energy depends (inversely quadratically) on the magnitude of the magnetic field in the neighborhood of the black hole, essentially that on the inner accretion disk.  The developing supernova explosion disrupts the accretion disk; this removes the magnetic fields anchored in the disk, and self-limits the energy the Blandford-Znajek mechanism can deliver."}  This, together with the total rate of creations of binaries of our type of $3\times10^{-4}$galaxy$^{-1}$yr$^{-1}$ estimated by \citet{Bro00} will be shown in section~\ref{sec-Pop} to reproduce the population of subluminous bursts in nearby galaxies.  It is clear that they are subluminous, at least in the cases of Nova Sco and Il Lupi because their Kerr parameters measured by the Smithsonian-Harvard coalition are indistinguishable from the natal predicted $a_\star=0.8$ within observational errors.  The disruption of the black hole disk will be discussed in detail in Appendix~\ref{app2}.

Our evolution of black hole binaries in our Galaxy might appear to be irrelevant for the long (high luminosity) $\gamma$-ray bursts because \citet{Fru06} show that these come chiefly from low metallicity, very massive stars in galaxies of limited chemical evolution, quite unlike our Milky Way.  However, we can construct a quantitative theory of the rotational energies of the black holes which power the central engine for the GRBs and Hypernovae in our Galaxy because we can calculate the black hole Kerr parameters.  Having this quantitative theory it is straightforward to apply it to explosions in low metallicity galaxies, the high luminosities resulting because the donors are more massive than those in our Galaxy and can accept more rotational energy from the binary than the high metallicity binaries with low mass donors, although there may be some overlap of the high metallicity stars with the same mass donor as the low metallicity stars, as we shall discuss.  We predict that this will be so with XTE J1550-564 (V383 Normae), in which the Kerr parameter will be measured (J. McClintock, private communication) so that with our calculation presented here we will be able to obtain the energy used up in the explosion which should be nearly that of cosmological (high luminosity) GRBs.

Recently the eclipsing massive black hole binary X$-$7 has been discovered in the nearby Spiral Galaxy Messier 33\footnote{See also \citet{Bul07}.} \citep{Oro07}.  Since the metallicity is $\sim0.1$ solar, we believe it to mimic low metallicity stars which are more massive than the Galactic ones.  The donor has mass $68.5\msun$ now, possibly $\sim80\msun$ earlier. We expect that this system may have gone through a dark explosion due to the high donor mass, which implies a low rotational energy as we discuss in next section.


\subsection{Evolution of Cyg X$-$1, V4641 Sgr and GRS 1915$+$105}\label{sec-slGRBs}

A highly relevant discussion for the V4641 Sgr evolution, which will be a template of an earlier GRS 1915$+$105 development was given by \citet{Pod03}, for Cyg X$-$1.  They discussed the latter as if the donor and black hole were very nearly equal in mass which we shall show will happen in the future for Cyg X$-$1, although its donor is now $\sim18\msun$ and black hole $\sim10\msun$.  The donor could have been substantially more massive when the black hole was born after common envelope evolution and lost mass by wind.

The donor in \citet{Pod03} had a stellar wind of $3\times 10^{-6} \msun$~yr$^{-1}$ \citep{Her95} throughout the evolution. Once the mass of the donor became reduced to a mass comparable to the mass of the black hole, the donor established thermal equilibrium and filled its Roche Lobe transferring mass at the rate of $4\times10^{-3}\msun$yr$^{-1}$. Because of the continuing wind loss the donor shrank significantly below its Roche Lobe and the system widened. The donor started to expand again after it had exhausted all of the hydrogen in the core and filled its Roche Lobe a second time. In this phase the mass
transfer reached a second peak of $\sim 4\times 10^{-4}\msun$~yr$^{-1}$, where mass transfer was driven by the
evolution of the H-burning shell.

The most interesting feature of this calculation was that the system became detached after the first initial time scale because of the stellar wind from the donor. Since the donor is close to filling its Roche Lobe, such a wind may be focussed towards the accreting black hole, as is inferred from the tomographic analysis of the mass flow in Cyg X-1 by \citet{Sow98}.

\citet{Sow98} decompose the stellar wind of the supergiant into two moments, one representing the approximately spherically symmetrical part of the wind and the second representing the focussed enhancement of wind density in the direction of the black hole.  The latter component of the wind transferred mass in an essentially conservative way (although the former would bring about mass loss).  We shall use the wind in both, V4641 Sgr and GRS 1915$+$105, to accrete matter from the donor to the black hole later on.  Note that the wind, transferring matter at hypercritical (much greater than Eddington) rate basically shuts off the initial Roche Lobe overflow, because it can transfer mass sufficiently by itself.  The second period of Roche Lobe overflow transfer is driven by the evolution of the H-burning shell; i.e., by the secondary star going red giant.

\citet{Pod03} say that ``irrespective of whether this particular model is applicable to Cyg X$-$1, the calculation... illustrates that it is generally more likely to observe a high-mass black-hole X-ray binary in the relatively long-lived wind mass-transfer phase following the initial thermal timescale phase which only lasts a few $10^4$ yr.  In this example, the wind phase lasts a few $10^5$ yr, but it could last as long as a few $10^6$ yr if the secondary were initially less evolved."

We believe the above scenario to apply not only to Cyg X$-$1, V4641 Sgr and GRS 1915$+$105, but also to LMC X$-$3 and M33 X$-$7, binaries with donors more massive than the black-hole companion.  We find that these all had dark explosions, as we develop below, basically because the donor had too high a mass at the time of the explosion to give an energetic GRB.

We suggest that the wind in these cases may resemble the tidal stream of \citet{Blo91}.  In the two-dimension system studied by these authors, they show that when $D/R_\star$ becomes less than $\sim2$, where $D$ is the distance between O-star and black hole and $R_\star$ is the radius of the primary (or accreting) star, the tidally-enhanced-wind accretion exceeds Bondi-Hoyle accretion, a factor that increases to several as $D/R_\star$ decreases.

We are thus now able, given the evolution of \citet{Pod03}, to describe in detail the crude mass transfer used in the evolution of GRS 1915$+$105 by \citet{Lee02}, as conservative mass transfer in wind.  We disagree, however, with the procedure of \citet{Pod03} to limit the accretion to Eddington.  Obviously there is no surface in the black hole onto which the binding energy of the infalling matter can be accreted and then furnish photons etc. which will be emitted so as to impede the other infalling matter.  In fact, once the matter goes across the event horizon it can no longer influence the matter that has not.  In the case of 1915$+$105, \citet{Lee02} found the average mass transfer rate to be $\dot{M}\sim10^{-5}\msun$yr$^{-1}$, about 200 times Eddington (see \citet{Bet03} p.355).

\begin{figure}
\begin{center}
\epsfig{file=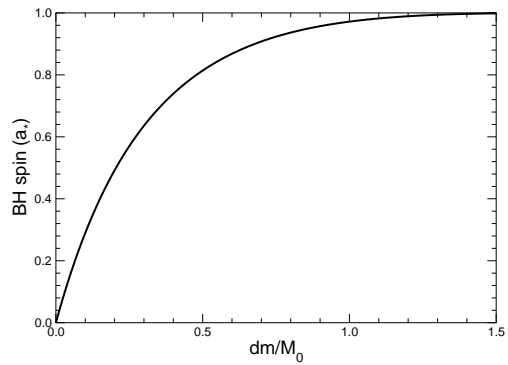,width=3in}
\end{center}
\caption{Spinning up of black holes. Black hole spin $a_\star$ is
given in units of [$GM/c^2$] and $\delta m$ is the total rest mass
of the accreted material. Note that $M_0$ is the mass of the
non-rotating initial black hole. Here we assumed that the last
stable orbit corresponds to the marginally stable radius \citet{Bro00}.
} \label{fig-a}
\end{figure}

None of the three binaries we consider had appreciable natal $a_{\star}$'s because of the high masses of the donors (secondary stars).  Thus, we can see that the high Kerr parameter must come from mass accretion (see Fig.~\ref{fig-a}) and that the accreted mass must be slightly greater than the natal mass in order to have
such a high Kerr parameter for GRS 1915$+$105.  Taking the final black hole mass to be $14\msun$ we need a natal mass of $\sim6\msun$.

If V4641 Sgr is to be a template for GRS 1915$+$105, then its black hole should have had a mass of $\sim6\msun$ when born, which would have evolved to its present value of $9.61\msun$ by accretion by wind from the donor.  Thus, the present $6.51\msun$ donor and $9.61\msun$ black hole will have essentially interchanged masses from the time of the birth of the black hole until present, through mass exchange from the donor to the black hole.  As outlined in LBW, V4641 Sgr will transfer by wind another $4.6\msun$ from donor to black hole to reach the present GRS 1915$+$105 with black hole mass $14\msun$.  Thus, the total mass accreted by GRS 1915$+$105 is estimated to be $\sim8\msun$.  The companion mass evolved in this way is $1.93\msun$, somewhat larger than the \citet{Har04} mass of $0.81\pm0.51\msun$.  Some mass is, however, lost in jets, which is not taken into account in our approximation of conservative mass transfer.

Our evolution of the black hole birth is similar to the second evolution version of \citet{Sad06} which would give the right chemical abundances to the secondary star in V4641 Sgr.  These investigators suggested that the black hole mass at birth was $7.2\msun$, and to obtain the right surface abundances they proposed that the explosion was a dark one; i.e., one of low energy.  \citet{Mir03} suggest that the explosion in Cyg X$-$1 was a dark explosion, much less energetic than the one in Nova Sco.  We shall see that the explosions are dark because the donors (secondary stars) have high masses.

\subsection{Energetics for Gamma Ray Bursters and Hypernovae}\label{sec-GRBs}

We will construct spin energies for the transient sources in the next section, but we want to make some general comments here.  The question with GRBs is whether there is enough angular momentum to power the GRB and Hypernova explosion.  In the case of the widely accepted theory, Woosley's Collapsars, this question is unanswered, although one may take the point of view that we observe GRBs and hypernova explosions, so there must be enough angular momentum, which is an integral part of the mechanism. None the less, \citet{Heg03} say ``when recent estimates of magnetic torques \citep{Spr02} are added, however, the evolved cores spin an order of magnitude slower. This is still more angular momentum than observed in young pulsars, but too slow for the collapsar model for gamma-ray bursts."  Furthermore, the usual scenarios for the interactions of Wolf-Rayets with other stars is that they slow down the rotation.

The above arguments in support of the binary model which was used in \citet{Bro00}, were made in \citet{Bro07}, and applied to Galactic Transient sources.

The Hypernova formed in 1998bw had $\sim3\times10^{52}$ergs in energy \citep{Iwa98}.  In addition, the jet formation in the GRB requires lifting all of the matter out of the way of the jet.  \citet{Mac00} estimates that this costs $\sim10^{52}$ergs in kinetic energy.  At early times the thermal and kinetic energies in supernova explosions are roughly equal, satisfying equipartition.  We believe this to be at least roughly true in our explosions here, so that $\sim6\times10^{52}$ergs would be needed for 980425/SN1998bw and possibly more\footnote{In fact, literally, his estimate of $(0.01-0.1)\msun$c$^2$ would be $(10^{52}-10^{53})$ergs.  Note that this energy is ``invisible", and is not normally included in estimates of GRB energies.}, because \citet{Mac00} describes 980425 as a ``smothered" explosion.

As noted in Appendix of Brown, Lee and Moreno M\'endez, the Blandford-Znajek efficiency drops substantially as the Kerr parameter decreases below $a_\star\sim0.5$.  Thus, the available rotational energy will decrease rapidly with increasing donor mass.  In Appendix~\ref{app2} we will estimate that the highest rotational energy of $\sim6\times10^{52}$ergs can be accepted by a binary with a $5\msun$ donor.


\section{Donor Mass and Black Hole Spin Anti-Correlation}\label{sec-LMD2HAM}

\subsection{Mass - Period Relation}

Using
\be
M_{He}=0.08(M_{Giant}/\msun)^{1.45}\msun
\label{eq:MHe}
\ee
LBW found that following common envelope evolution
\be
a_f=\left(\frac{M_d}{\msun}\right)
\left(\frac{M_{Giant}}{\msun} \right)^{-0.55}a_i\label{eq:af}.
\ee
Here $a_f$ is the final separation of the He star which remains from the giant following the strip off of its H envelope, and $a_i$ is its initial separation.  It has inherited the angular momentum of the He star and is tidally locked with the donor.  The giant masses found by LBW were all about $30\msun$.  From Kepler we have the preexplosion period
\be
\frac{\rm days}{P_b}=\left(\frac{4.2\rsun}{a_f}\right)^{3/2}
\left(\frac{M_d+M_{He}}{\msun}\right)^{1/2}. \label{eq:Kepler}
\ee

\begin{table}
\begin{center}
\begin{tabular}{|c|c|c|c|c|}
\hline
 Name           &$M_{BH}$ &  $M_{d}$  &$a_{\star}$&   $E_{\rm BZ}$    \\
                &[$\msun$]& [$\msun$] &           &    [$10^{51}$ ergs]      \\
\hline
\hline
\multicolumn{5}{|c|}{AML: with main sequence companion} \\
\hline
J1118$+$480     & $\sim 5$ & $<1$     &   $0.8$   &  $\sim 430$ \\
Vel 93          & $\sim 5 $ & $<1$    &   $0.8$   &  $\sim 430$ \\
J0422$+$32      & 6$-$7  &  $<1$    &  0.8 & $500\sim 600$ \\
1859$+$226      & 6$-$7  &  $<1$    &  0.8 & $500\sim 600$ \\
GS1124          & $6-7$  &  $<1$    &  0.8 & $500\sim 600$ \\
H1705           & $6-7$  &  $<1$    &  0.8 & $500\sim 600$ \\
A0620$-$003     & $\sim 10$ & $<1$   & 0.6  &  $\sim 440$ \\
GS2000$+$251    & $\sim 10$ & $<1$   & 0.6  &  $\sim 440$ \\
\hline
\hline
\multicolumn{5}{|c|}{Nu: with evolved companion} \\
\hline
GRO J1655$-$40  & $\sim5$ &  1$-$2    &   $0.8$   &   $\sim 430$ \\
4U 1543$-$47    & $\sim5$ &  1$-$2    &   $0.8$   &   $\sim 430$ \\
XTE J1550$-$564 &$\sim10$ &  1$-$2    &   $0.5$   &   $\sim 300$ \\
GS 2023$+$338   &$\sim10$ &  1$-$2    &   $0.5$   &   $\sim 300$ \\
XTE J1819$-$254 &  6$-$7  & $\sim10$  &   0.2                   &  $10\sim 12$ \\
GRS 1915$+$105  &  6$-$7  & $\sim10$  &  0.2 ($>0.98^\dagger$)  &  $10\sim 12$ \\
Cyg X$-$1       &  6$-$7  &$\gtrsim30$&   0.15                  &  $ 5 \sim 6$ \\
\hline
\end{tabular}
\end{center}
\caption{Parameters at the time of black hole formation.
$E_{\rm BZ}$ is the rotational energy which can be extracted via Blandford-Znajek mechanism with optimal efficiency $\epsilon_\Omega=1/2$ (see Appendix~\ref{app1}).
The AML (Angular Momentum Loss) binaries lose energy by gravitational waves, shortening the orbital period whereas the Nu (Nuclear Evolution) binaries will experience mass loss from the donor star to the higher mass black hole and, therefore, move to longer orbital periods. $^\dagger$ Kerr parameter is the present one.
}\label{tab-a}
\end{table}

From Fig.~\ref{FIG12} (Fig.12 of LBW) one sees that the Kerr parameter increases sharply as the period of the binary $P_b$ decreases.  From eq.~(\ref{eq:af}) we see that $a_f$ is proportional to $M_d$, the donor mass, and from
eq.~(\ref{eq:Kepler}) that $P_b$ is proportional to $a_f^{3/2}$.  Estimated Kerr parameters for Galactic sources are summarized in Table~\ref{tab-a}, which makes the dependence on the donor clear for the galactic sources.

In fact, it would seem \'a priori more natural to have an $\sim10\msun$ donor in the binary with a $30\msun$ giant than a $2.5\msun$ donor and $30\msun$ giant as in Nova Scorpii.  We shall see further on in this section that Case C mass transfer requires that the initial $a_i$ is very large, $\sim2000\rsun$ for a $30\msun$ giant.  The increased gravitational binding between donor and the He-star left from the giant after being stripped of hydrogen must furnish the energy, modulo the efficiency $\lambda\alpha_{ce}$, to remove the hydrogen envelope.  The latter decreases inversely with the radius of the giant, so that when the radius is large, the envelope can be removed by a low-mass companion.  Combining eqs.~(\ref{eq:af}) and~(\ref{eq:Kepler}) we have
\be
\frac{\rm days}{P_b}= \left(\frac{4.2\rsun/a_i}{M_d/\msun}\right)^{3/2} \left(\frac{M_d+M_{He}}{\msun}\right)^{1/2} \left(\frac{M_{giant}}{\msun}\right)^{0.83} \label{eq:afKepler}
\ee
so that $P_b\propto M_d^{3/2}$ for $M_d \ll M_{He}$ and $P_b\propto M_d$ for higher donor masses.  As we develop later, the distances $a_i$ at which mass transfer begins depend very little on donor mass so we can use eq.~(\ref{eq:afKepler}) in order to scale from one donor mass to another.  LBW found all giants they needed for the transient sources had mass $\sim30\msun$, and, therfore, $M_{He}\sim 11 \msun$, substantially larger than the donor masses for the most energetic GRBs.

For higher mass donors mass loss in the explosion can be neglected, giving pre and post-explosion periods which are nearly the same.

\subsection{Explosion Energies of Galactic Black Hole Binaries}

In supernova explosions at short times the kinetic and thermal energies are equal, following equipartition.  We are close to finding this to be true for GRBs and Hypernovae.  The kinetic energy, which is an order of magnitude or more greater than the GRB energy \citep{Mac00}, results from the ram pressure which is needed to clear the way for the jet which initiates the GRB, as just described.  We shall assume the kinetic energy to be the same as the thermal energy, which is more easily measured from the observations.  Thus, we assume that twice the hypernova energy is needed to power the explosion.  On the other hand, the efficiency $\epsilon_\Omega$ in depositing the energy in the perturbative region is usually taken to be $1/2$, the optimum value.  This optimum value is obtained by impedance matching as in ordinary electric circuits.

The Blandford-Znajek energy is deposited in a fireball in the perturbative region.  \citet{Pac86} and \citet{Goo86} have shown that in order to power a GRB all that is necessary is for the fireball to have enough energy so that the temperature is well above the pair production temperature; i.e., $T>1$MeV.  Then the GRB and the afterglow will follow, just from having as source the localized hot fireball.

The estimated explosion energies of Galactic sources are summarized in Table~\ref{tab-a}.
We claim that ours is the first quantitative calculation of the explosions giving rise to GRBs and Hypernovae, in the sense that we calculate the energy that is supplied in the form of angular momentum energy.  What the central engine does with this energy is another matter.  We cannot calculate how much of the energy is accepted by the accretion disk in detail, because we cannot calculate analytically the Rayleigh-Taylor instability in magnetic field coupling to the accretion disk of the black hole (but see Appendix~\ref{app2}).

We do know from observations that accretion disks have been formed; the Kerr parameters have been measured by relativistic Doppler effects as the matter goes into the black hole.  We also know that jets are launched sporadically from the soft X-ray transient sources in our Galaxy.

The rotational energy that is not accepted by the black hole does, however, remain in the binary and appears later in the Kerr parameter of the black hole.  Thus far, in Nova Sco and Il Lupi a tiny part of the available rotation energy
was accepted by the central engine, so little that the final rotational energy could not be discriminated from our calculated initial energy, because the uncertainty in the measurement of Kerr parameters was the same order of magnitude as the explosion energy.

\begin{figure}
\begin{center}
\epsfig{file=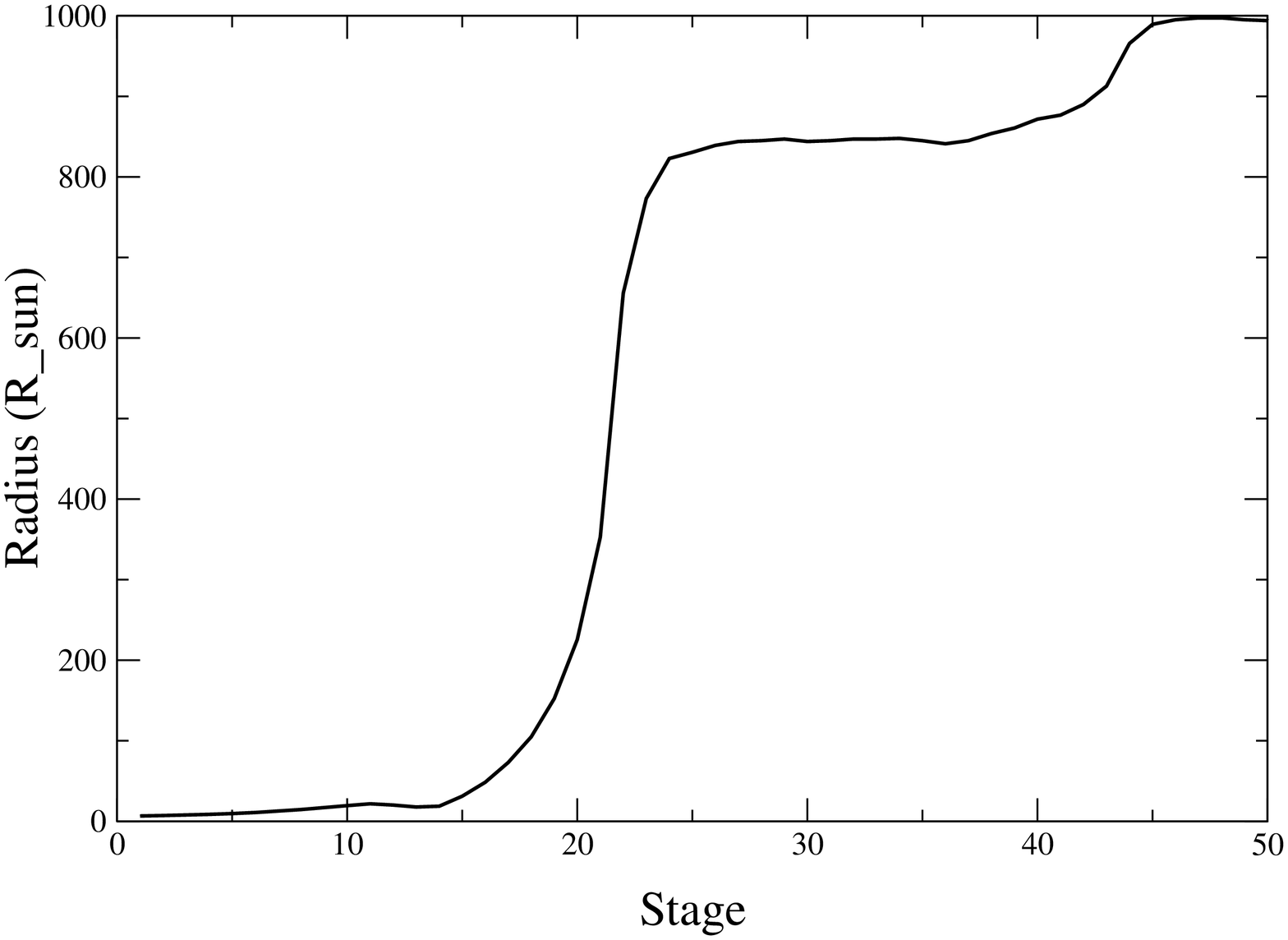,width=3in}
\end{center}
\caption{Radial expansion of the 25 ZAMS mass star, obtained by
LBW from modifying \citet{Sch92}. Helium core burning
ends at stage 43.}\label{fig2}
\end{figure}

We can understand why AMLs and Nova Sco, Il Lupi, XTE J1550$-$564 and GS 2023$+$338 among Nus have such rotational energies from Case C mass transfer.  The lower the companion mass, the greater the radius $R_{sg}$ that the supergiant must reach before its Roche lobe meets the companion.  Given giants such as shown in Fig.~\ref{fig2}, the typical separation distance between giant and companion is $\sim(1200-1300)\rsun$, higher (because of the Roche lobe of the companion star) than the $1000\rsun$ the giant radius reaches.  The binding energy of the supergiant envelope goes as $R^{-1}_{sg}$ and at such a large distance it can be removed by the change in binding energy of an $\sim1\msun$ donor, spiralling in from $\sim1500\rsun$ to $\sim5\rsun$.  A higher mass donor would end up further out, since it would not have to spiral in so close in order to release enough binding energy.  Thus low mass $\sim(1-2)\msun$ companions can naturally deposit their increase in gravitational binding energy in removing the high $R_{sg}$ envelopes.  A detailed discussion of these matters is given in \citet{Bro99}.  Thus Case C mass transfer naturally gives the ultra-high rotational energies of the binaries with low-mass donors discussed as relics of GRB and hypernova explosions in our Galaxy.

{\bf Cyg X-1:} One can see that there was essentially no mass loss in the Cyg X$-$1 explosion because the space velocity of Cyg X$-$1 relative to Cyg OB3 in the cluster of O-stars of $(9\pm2)$km s$^{-2}$ is typical of the random velocities of stars in expanding associations \citep{Mir03}.  In comparison, Nova Sco had a very strong explosion from the fact that its space velocity after explosion is $(112\pm18)$km s$^{-1}$, although only a small part of the available energy is used up in the system velocity.  Our reconstruction of the Cyg X$-$1 evolution had the explosion take place when the black hole mass was about $7\msun$, the $3\msun$ less than the black hole mass is today having accreted from the donor (secondary star).  The mass transfer from donor to black hole is nonconservative because of the higher mass of the donor.  The donor would have been substantially more massive than it is today at the time of the explosion, at least $\sim30\msun$.  For a $30\msun$ donor we obtain the maximum available energy $<10^{52}$ergs.

\citet{Mir03} believe that Cyg X-1 had a dark explosion.  As discussed in Appendix~\ref{app1}, the efficiency $\epsilon_\Omega$ can be taken to be $0.5$ for the higher $a_\star$, say $a_\star>0.5$, but it decreases for small $a_\star$, so that for $a_\star=0.15$ we calculate it to be $0.15$ and for $a_\star=0.2$ we calculate $\epsilon_\Omega=0.2$.  The $(5-6)\times10^{51}$ergs is clearly not enough for an explosion in Cyg X$-$1.

{\bf Binaries in low metallicity galaxies:} As discussed in a previous section, \citet{Oro07} have recently measured the extra Galactic M33 X$-$7 in a neighborhood where the metallicity is $\sim0.1$ solar.  Following our prediction that the donors of low-metallicity galaxies are generally more massive than those of our Galaxy, the now $68\msun$ ($\sim80\msun$ at the time of common envelope evolution) is much more massive.  In fact, it is too massive in that with the small Kerr parameter we estimate to be $a_\star=0.12$; it probably went through a dark explosion.

Far from being irrelevant, as \citet{Fru06} imply (because of their lack of dynamics), the measurements of the Harvard-Smithsonian group teach us how to calculate the energies of GRB and Hypernova explosions.  Having a dynamical theory, we can easily extend it to low metallicity galaxies, by increasing the donor masses.

\subsection{Subluminous Bursts}
\label{sec-Pop}

All of the GRBs in our binary model come from the same mechanism, but their angular momentum energy will be decided by the mass of their donor.  Our mechanism suggests, however, that the binaries are usually left spinning with the measured Kerr parameter.  There must be a ``Goldilocks" scenario for the energy needed to power a high-luminosity GRB, neither too big nor too small.  Galactically, with the $(1-2.5)\msun$ low mass donors, the energy is clearly too large.  We know this because the calculated initial Kerr parameters were essentially the same as those found by \citet{Sha06}; thus, very little of the energy had been used up in the explosion.  On the other hand we have M33 X$-$7, which we will discuss in more detail later, which had a donor of $\sim80\msun$, with Kerr parameter $a_\star=0.12$ which probably went into a dark explosion, like Cyg X$-$1.  So we have bracketed (but rather widely) the luminous explosions.

Initially in supernova explosions, the kinetic energy, the main part of which results from clearing out the matter in the way of the jet that accompanies the GRB, is about equal to the thermal energy of the hypernova.  \citet{Mac00} finds this to be at least approximately true and it would follow from equipartition of energy.  We should be able to connect GRS 980425 with our galactic GRBs because of its high metallicity, nearly solar \citep{Sol05}.  (The galaxy of GRB 980425 is incorrectly put in the class of low-metallicity by \citet{Sta06} and by \citet{Heu07}.)

Note that the high luminosity GRBs turn out to be only a small fraction of the total number, even if we use a beaming factor of 100 for them.  Thus, they must be formed in very special circumstances (see Appendix~\ref{app2}).

The question is whether there are enough transient sources to supply subluminous GRBs in nearby galaxies.  \citet{Bro00} estimated that in our Galaxy the total rate of creation of the transient source binaries was $\sim3\times10^{-4}$galaxy$^{-1}$yr$^{-1}$.  Given $10^5$ galaxies within $200$Mpc this number translates into $3,750$Gpc$^{-3}$yr$^{-1}$.  \citet{Lia07} find a beaming factor typically less than 14; such a beaming factor would reduce our number to $268$Gpc$^{-3}$yr$^{-1}$, essentially that of \citet{Lia07} of $\sim325^{+352}_{-177}$Gpc$^{-3}$yr$^{-1}$.  This is much higher than their rate of high-luminosity GRBs of $1.12^{+0.43}_{-0.20}$Gpc$^{-1}$yr$^{-1}$.  The usual beaming factor for the high luminosity bursts is $\sim100$.  Even with such a large factor, the high luminosity GRBs are estimated to be much less in number, by a factor of $\sim40$, than the subluminous ones.  Although one should add the Woosley Collapsar rate to our binary rate, we have enough binaries to account for all of the bursts. \citet{Woo06} estimate that $\sim1\%$ of the stars above $10\msun$ can, under certain circumstances, retain enough angular momentum to make GRBs.

The effect of cutting down the wind losses Galactically gave a hint about how the rotational energy in the binaries could be decreased so as to be in the ballpark needed for high lumiosity GRBs.  The winds are particularly high because of Galactic metallicity.  Low metallicity stars have much less wind.  In general the stars are more massive than Galactic ones, which we believe has the effect of scaling up all of the Galactic masses.
We pursue the question of cosmological GRBs and their abundances in Appendix~\ref{app2}.

\subsection{A General Discussion of Black Hole masses}
\label{BHMass}

If one accepts the \citet{Sch92} numbers literally, then Case C mass transfer is actually limited to a narrow interval of ZAMS masses about $20\msun$, $\sim(19-22)\msun$ as found by \citet{Por97}.  This is because the binary orbit widens with mass loss of the supergiant so that in order to initiate mass transfer only after helium burning the supergiant has to expand sufficiently that this widening of the orbit is compensated for.  A graphic display of this is shown in Fig. 1 of \citet{Bro01b}.

LBW realized that in order to reproduce black holes from the interval of ZAMS masses $(18-35)\msun$, necessary for their evolution of transient sources, they had to cut down the wind losses in the red giant stage (by hand -see LBW Fig. 3).  This was clearly necessary because \citet{Bro01a} had shown that high mass X-ray black hole binaries could be evolved with the black holes coming from ZAMS masses $(18-35)\msun$ provided Case C mass transfer was used.
It may be that the donor mass for high luminosity GRBs has to be higher than $5\msun$.
We do not yet know how rapidly the binaries are left rotating after the explosion.  Measurements of black hole binaries with donor masses $\sim(10-20)\msun$ would be very helpful.

In LBW it seemed strange that the giant progenitors of the black-hole binaries Nova Sco, Il lupi and GRS 1915$+$105 all came from $(30-33)\msun$ giants, whereas black holes were formed from ZAMS mass $(18-35)\msun$ in Case C mass transfer in our Galaxy, and the lower mass black holes are certainly more copius than $(30-33)\msun$ ones.

In the case of Nova Sco and Il Lupi the explosion was so energetic that the black hole of $5.5\msun$ was only about half of the progenitor He star mass, i.e. the explosion was so violent that nearly half of the mass of the system was lost in the explosion; a loss of half or more resulting in system breakup.  GRS 1915$+$105 and Cyg X$-$1 did have black holes of $(6-7)\msun$, with little mass loss in the explosion, which came from $(20-22)\msun$ progenitors (our evolution of GRS 1915$+$105 in the present paper is an improvement over that in LBW).  Thus, the black holes in Galactic soft X-ray transient sources do really come from a wide range of ZAMS mass progenitors.

\section{Conclusions}\label{sec-Conclusions}

Our theory of GRBs and hypernova explosions was developed in \citet{Bro00} and is essentially unchanged.  In the meantime we learned in \citet{Lee02} how to calculate the Kerr parameters of the black holes, essentially through an understanding of the tidal locking.    Our Kerr parameters have been checked in our Galaxy by the measurement of the Kerr parameters of Nova Sco and Il Lupi by the Smithsonian-Harvard coalition.  Both \citet{Pac86} and \citet{Goo86} have shown that when sufficient energy has been delivered to the fireball (so that the temperature is above the pair-production threshold) the GRB and Hypernova explosions follow and the afterglow is that as observed.  In this sense the production of energy and the explosion decouple, but the latter follows from the former once sufficient energy is furnished.  In this sense we have a complete calculable theory of GRBs and Hypernovae.

The GRB and Hypernova explosions are just those of the Collapsar model of Woosley, but with an important improvement; namely, any required amount of rotational energy is obtained from the tidal spin up of the black-hole progenitor by the donor.  The donor then, after furnishing the angular momentum, acts as a passive witness to the explosion, but can show some detail of the latter in the chiefly alpha-particle nuclei which it accretes.
The magnetic field lines threading the disk of the black hole are well placed to power the central engine in the Blandford-Znajek mechanism.  The jet formation and hypernova explosion are powered just as in the \citet{Mac99} Collapsar.  We check by population synthesis that our binaries are sufficient in number to reproduce all GRBs.

The great advantage that the soft X-ray transient sources in our Galaxy have is that their properties can be studied in detail.  They are, however, a special class because of the high metallicity in our Galaxy.
None the less, it is easy to extend our galactic description to one of low-metallicity galaxies, because the angular momentum energy is determined by the mass of the donor.  Donors in low-metallicity galaxies tend to be more massive than in high-metallicity ones, furnishing a lower rotational energy.

We find that the subluminous GRBs come from two sources: 1) The Galactic metallicity with low-mass donors, where the magnetic field coupling to the black hole disk is so high that it dismantles the central engine before much angular momentum energy can be delivered.  Nova Sco and Il Lupi are excellent examples of these, in that we have shown that only a tiny part of the angular momentum energy was used up in the explosion of these.  2) Binaries with low metallicity donors which are massive, going up to the $80\msun$ donor in M33 X$-$7.  Somewhere in between these extremes the binaries will have the rotational energies of the cosmological GRBs.

We estimate the high luminosity GRBs to come from binaries with donor masses $\sim5\msun$, but this is uncertain until the Kerr parameters of binaries such as XTE J1550$-$564 are measured.  With such a Kerr parameter in hand, we can subtract the rotational energy left in the binary from the preexplosion energy which we calculate (see Table~\ref{tab-a}).  This will tell us the energy of the explosion.  In the cases of Nova Sco and Il Lupi, the energy used up in the explosion was tiny compared with the initial rotational energy, but this must change as the initial rotational energy decreases, and black hole mass increases.

We have shown that Galactically there are 12 relics of GRBs and hypernova explosions, and that 3 (XTE J1819$-$254, GRS 1915$+$105, Cyg X$-$1) might have gone through a low-energy dark explosion\footnote{Called ``smothered" explosion by \citet{Mac00}.}, although the first two of these may have gone through GRBs and hypernova explosions. So we believe that the soft X-ray black hole binaries are the major sources for the subluminous GRBs.

We suggest that the luminous cosmological GRBs result from a ``Goldilocks" phenomenon, and that only the binaries with donor masses $\sim 5\msun$ have enough energy for cosmological GRBs.


\begin{acknowledgments}

We would like to thank Jeff McClintock for many useful discussions. We believe that the Smithsonian-Harvard group and
collaborators, who did not know of our predictions for the $a_\star$'s of Nova Scorpii and IL Lupi have opened up an exciting new field of activity. G.E.B. was supported in part by the US Department of Energy under Grant No. DE-FG02-88ER40388.
CHL was supported by National Nuclear R\&D Program(M20808740002) of MEST/KOSEF .
\end{acknowledgments}



\appendix

\section{Appendix A:  Blandford-Znajek Mechanism}\label{app1}

\begin{figure}
\centerline{\epsfig{file=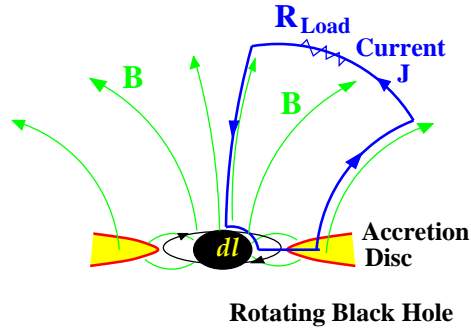,height=1.7in}}
\caption{The black hole in rotation about the accretion disk, formed by what is left of the He star. A wire loop can be drawn, coming down a field line from the top (this particular field line is anchored in the black hole.) to the north pole of the black hole. The black hole has a surface conductivity, so the wire can be extended from the north pole of the black hole to the equator and further extended into the (highly ionized) accretion disk, in which the magnetic field lines are frozen. The wire can be continued on up out of the accretion disk along a field line and than connects back up to form a loop. As this wire loop rotates, it generates an electromagnetic force, by Faraday's law, sending electromagnetic radiation in the Poynting vector up the vertical axis. The region shown has condensates of charge designed to allow free rotation of
the black hole, although the black hole when formed rotates much more rapidly than the accretion disk, since the compact object angular momentum must be conserved as the inner part of the He star drops into the black hole. In trying to spin the accretion disk up, the rotation engendered by the field lines is converted to heat by viscosity, the resulting hypernova explosion taking place in a viscous time scale. The gamma ray burst is fueled by the deposition of Poynting vector energy which is sent up the rotational axis into a fireball.}
\label{bzcirc}
\end{figure}

In Fig.~\ref{bzcirc} we show that a rotating black hole operates like a generator of electricity; this is the Blandford-Znajek \citep{Bla77} mechanism, which is summarized in the caption. We rely on the relatively complete review by \citet{Lee00}.

The rotational energy of a black hole with angular momentum $J$ is a fraction of the black hole mass energy
\be
E_{\rm rot} = f(a_\star) M_{\rm BH} c^2, \label{eqa1}
\ee
where
\be f(a_\star) = 1 - \sqrt{\frac 12 \left(1+\sqrt{1-a_\star^2}\right)}.
\ee
For a maximally rotating black hole ($a_\star=1$) f=0.29. In Blandford-Znajek mechanism the efficiency of extracting the rotational energy is determined by the ratio between the angular velocities of the black hole $\Omega_H$ and the magnetic field velocity $\Omega_F$,
\be
\epsilon_\Omega = \Omega_F/\Omega_H.
\ee
For optimal energy extraction $\epsilon_\Omega \simeq 0.5$. This just corresponds to impedance matching between that in the generator (Fig.~\ref{bzcirc}) and that in the perturbative region where the energy is delivered. One can have the analytical expression for the energy extracted
\be
E_{BZ} = 1.8\times10^{54}\epsilon_{\Omega}f(a_{\star})\frac{M_{BH}}{\msun}{\rm erg}. \label{EBZ}
\ee
The rest of the rotational energy is dissipated into the black hole, increasing the entropy or equivalently irreducible mass.

Although the use of $\epsilon_\Omega=0.5$ is close to the actual efficiency for high $a_\star$, it decreases with $a_\star$ from $0.69$ at $a_\star\sim0.8$ to $0.46$ at $a_\star\sim0.4$ (\citep{Bro00}, see $\Omega_{disk}/\Omega_H$, $r=r_{ms}(\tilde{a})$ where $r_{ms}$ is the marginally stable radius).  There would be a further decrease of $\gtrsim50\%$ to the $a_\star$ of $0.15$ of Cyg X$-$1 and M33 X$-$7.  This low efficiency virtually ensures that these two binaries will have gone through dark explosions.

The hypernova results from the magnetic field lines anchored in the black hole and extending through the accretion disk, which is highly ionized so the lines are frozen in it. When the He star falls into a black hole, the latter is so much smaller in radius that it has to rotate much faster than the progenitor He star so as to conserve angular momentum.

Initially large amount of energy up to $10^{52}$ ergs were attributed to gamma ray bursts. However, when correction was made for beaming the actual gamma ray burst energy is distributed about a ``mere" $\sim 10^{51}$ ergs \citep{Pir02}.  However, $\sim10^{52}$ergs are required to clear the way for the jet.  Hypernova explosions are usually modeled after the nearby supernova 1998bw. The hypernova by \citet{Nom00} that \citet{Isr99}  compared with Nova Scorpii had an energy of $3\times 10^{52}$ergs.


\section{Appendix B:  Dismantling the Accretion Disk by High Energy Input}\label{app2}

The amount of energy poured into the accretion disk of the black hole, and, therefore, also pressure is almost unfathomable., the $4.3\times10^{53}$ergs being $430$ times the energy of a strong supernova explosion, the latter being spread over a much larger volume than that of the accretion disk.  Also, $4.3\times10^{53}$ergs$\simeq\frac{1}{4}\msun c^2$.  Near the horizon of the black hole, the physical situation might become quite complicated \citep{Tho86}.  Field-line reconstruction might be common and lead to serious breakdowns in the freezing of the field to the plasma; and the field on the black hole sometimes might become so strong as to push it back off the black hole and into the disk (Rayleigh-Taylor Instability) concentrating the energy even more.  During the instability the magnetic field lines will be distributed randomly in ``globs", the large ones having eaten the small ones.  It seems reasonable that the Blandford-Znajek mechanism is dismantled.  Later, however, conservation laws demand that the angular momentum not used up in the GRB and hypernova explosion be reconstituted in the Kerr parameter of the black hole.  The radius of the (Kerr) black hole is
\begin{eqnarray*}
R=\frac{GM}{c^2}=
1.48\times10^{6}\frac{M}{10\msun}{\rm cm}.
\end{eqnarray*}

Given the above scenario of the very high magnetic couplings dismantling the disk by Rayleigh-Taylor instability, in Nova Sco, we wish to make a ``guestimate" of the same effect for XTE J1550$-$564, since its Kerr parameter is being measured by the Smithsonian-Harvard collaboration.

The black hole radius is proportional to $M_{BH}$.  The ratio of $M_{BH}(1550-564)/M_{BH}(1655-40)\simeq2$, actually more like $2.5$ because GRS 1655$-$40 has a Kerr black hole and in XTE J1550$-$564 the black hole is about halfway between Kerr and Schwarzschild.  Thus, the area of the last stable circular orbit is $\sim6$ times larger for XTE J1550$-$564.  Taking the field strength for Rayleigh-Taylor instability, to go as the inverse of the area, this means that its effect would be cut down by a factor of 6 in going from GRS 1655$-$40 to XTE J1550$-$564.  If our scenario that the magnetic field coupling is correct for Nova Sco, not much of the effect would be left in XTE J1550$-$564
which should accept most of the energy for a cosmological GRB.

The question then is, what is the latter?  We know that SN1998bw had a hypernova energy of $\sim30$bethe.  From the equipartition of energy, we would estimate the kinetic energy to clear out a path for the jet in the GRB to be about equal to the thermal energy, which is roughly true in \citet{Mac00}.  So the total energy would be $\sim60$bethes, which we know can be accepted by the binary.  \citet{Mac00} suggested that the accompanying GRB 980425 was ``smothered", so that may be a lower limit, although it is larger than estimates we have seen to date, so we choose it as the energy of a cosmological GRB.

Now, 60bethes is $\sim20\%$ of our estimated angular momentum energy for XTE J1550$-$564.  However, the decrease in Kerr parameter necessary to go from 300 to 240bethes is only $0.06$ from our natal $a_\star\sim0.5$, or $\sim12\%$, which requires an accurate measurement in Kerr parameter, but is none the less much larger than the $\sim6\%$ decrease estimated for Nova Sco.  (Note that for Schwarzschild black holes, the rotational energy goes approximately as $a_\star^2$.)  On the other hand, this may be the minimal difference between natal and present angular momentum energies because no other model leaves the system spinning so rapidly.  Thus, if no difference is conclusively seen because observational errors are $\sim12\%$ then this is also very interesting.

Suppose a decrease of more than $12\%$ is seen.  Then this means that we have underestimated the energy of the explosion, but our above estimates are as large as any proposed ones.  In any case, the possibility that the system, following the explosion, is left rotating is a new and interesting one.

We thus see that we can fit the \citet{Fru06} condition for no high luminosity GRBs in our high metallicity stars in it, because all of the donor masses of the GRBs that received the highest rotational energies had companion masses of $\sim(1-2)\msun$ and the rotational energy furnished to them was so great that the accretion disks were dismantled.  The result is that the GRBs were subluminous, like the vast majority of GRBs.  The $\sim6$ times larger surface area should be helpful in allowing XTE 1550$-$564 to accept the rotational energy, but the amount is still tremendous.  We do not have any donors of $\sim5\msun$ in our Galaxy, but such a donor would bring the energy down by $\sim1/5$, since it goes inversely with donor mass, to $\sim60$bethes, that we estimate for GRB 980425, which as noted has nearly Galactic metallicity.  Although GRB 980425 is essentially ``smothered" \citep{Mac00}, $60$bethes is the highest energy anyone has attributed to the GRB and Hypernova explosion.
Thus, our estimates would say that donors of $\sim5\msun$ would give the most luminous GRBs, and that XTE 1550$-$564 may or may not have been able to accept $\sim60$bethes, but since we view this as an upper limit, this binary should still be spinning furiously.

We should enter a proviso here.  Our main thesis is that the black hole binary must have a donor sufficiently massive to slow it down enough so that the black hole can accept the strong magnetic coupling through its accretion disk without the Rayleigh-Taylor instability entering.  By increasing the area of the accretion disk by a factor $\sim6$ in going from the $\sim5\msun$ black hole in Nova Sco, to the $\sim10\msun$ black hole in XTE 1550$-$564 the density of magnetic coupling is decreased by a factor $\sim6$.  However, the donor in XTE 1550$-$564 is only $\sim1.3\msun$, about the same size as in Nova Sco.  Therefore, the GRB in XTE 1550$-$564 may still have been subluminous, but less so than Nova Sco because of the larger disk area.

The donor masses at the time of common envelope evolution of XTE 1819$-$254 and GRS 1915$+$105 were $6\msun$.
However they have only $\sim1/3$ of the energy of the binary with $5\msun$ donor, $\sim(10-12)$bethes.
These black holes accrete a lot of matter, more than doubling the black hole mass in the case of GRS 1915$+$105 and, ultimately will double that mass in XTE 1819$-$254.
Such binaries with substantial mass exchange do not obey our simple scaling which is designed for natal angular momentum.  They must be evolved in detail.

We believe, that similar evolutionary arguments will apply to binaries with higher mass donors, and, anyway, the angular momentum energy will be cut down by the higher donor masses.  Thus we expect that only binaries with donor masses $\sim5\msun$ will give highly luminous GRBs with rotational energy $\sim60$bethes.

In a very rough estimate, using a flat distribution of donors with mass up to $80\msun$ we can estimate that the number of cosmological GRBs, those with high luminosity will be $\sim5/80$ of the total, not far from the ratio of GRBs with high luminosity to subluminous ones found by \citet{Lia07}.

In summary, the commonly accepted estimates of the explosion energies in GRBs are orders of magnitude less than the natal angular momentum energy in Nova Sco.  The measured Kerr parameter of $a_\star\sim0.8$ \citep{Sha06} has a present rotational energy indistinguishable from the natal one, within error bars.  However, the dismantling of the accretion disk from the very high magnetic couplings should be less in XTE J1550$-$564 and the rotational energy is somewhat less, the donor being about the mass of the He star in the Woosley Collapsar model, so we propose that XTE J1550$-$564 can accept substantial rotation energy but probably less than the energy of a cosmological GRB.  Measurement of this energy is being carried out by the Smithsonian-Harvard collaboration.

If our suggested scenario is confirmed, then this should be strong support for the
\citet{Bro00} binary scenario for GRBs.


\section{Appendix C:  Hypercritical Accretion}\label{app3}

\citet{Bro94} calculated analytically hypercritical spherical accretion onto compact objects and in particular for the fall back in SN 1987a and obtained the same result \citet{Hou91} obtained numerically.  Along this report we have used the evolution of Cyg X$-$1 by \citet{Pod03} where they find mass transfers from the secondary to the black hole as large as $10^{-4}\msun$ per year for a period on the order of $10^4$ years only to limit the accretion to Eddington's limit in the end, ejecting most of the transferred mass out of the binary.  We have used the \citet{Pod03} path in order to evolve not only Cyg X$-$1, but also V4641 Sgr and GRS 1915$+$105, except we believe a large fraction of this mass gets accreted hypercritically into the black hole.  Similarly, \citet{Mor08} show that M33 X$-$7 can only be evolved into its current state \citep{Liu08}, if hypercritical accretion takes place in this system.  Next we outline the calculation by \citet{Bro94}:

\citet{Bro94} obtain, for SN 1987a, an accretion rate of
\be
\dot{M}=1.15\times10^{22}{\rm g  s}^{-1}=1.81\times10^{-4}\msun {\rm yr}^{-1}, \label{eq:dotM}
\ee
which is the same order of magnitude as the one in \citet{Pod03}, so we simply follow their results.  The Eddington accretion rate is
\be
\dot{M}_{\rm Edd}=\frac{4\pi cR}{\kappa_{\rm es}}=5.92\times10^{-8}\msun{\rm yr}^{-1}, \label{eq:Edd}
\ee
where $R\simeq10^{6}$ cm is the radius of the compact object and $\kappa_{\rm es}$ is the opacity, which we take to be $\kappa_{\rm es}\simeq0.1$ cm$^2$ g$^{-1}$.  This is an estimate that applies over a range of temperature and density similar to that present here \citet{Che81}.  The Eddington luminosity $L_{Edd}$, the luminosity for which the pressure of outward traveling photons balances the inward gravity force on the material, is obtained from $L_{\rm Edd}=\dot{M}_{\rm Edd}c^2$.  If $\dot{M}$ exceeds $\dot{M}_{\rm Edd}$ then some of the accretion energy must be removed by means other than photons.  In the present case,
\be
\dot{m}\equiv\frac{\dot{M}}{\dot{M}_{\rm Edd}}=0.31\times10^4.
\ee
When $\dot{M}$ exceeds $\dot{M}_{\rm Edd}$ by so large a factor, the accretion rate is called hypercritical, and was considered by \citet{Blo86}.

\citet{Blo86} finds a trapping radius $r_{tr}$ such that photons within $r_{\rm tr}$ are advected inward with accreting matter faster than they can diffuse outward.  We follow his derivation in slightly modified form.  We start with the same type of equation as is used in deriving the Bondi $\dot{M}$:
\be
\dot{M}=4\pi r^2\rho v. \label{eq:Blondi}
\ee
Here, however, we consider any radius $r$ and take $v$ to be the free fall velocity, that is $v=(GM/r)^{1/2}$.  We divide equation~\ref{eq:Blondi} by equation~\ref{eq:Edd} to find
\be
\rho\kappa_{\rm es}=Rr_s^{-1/2}r^{-3/2}\dot{m},
\ee
where $r_{\rm s}=2GM/c^2$ is the Schwarzchild radius; $r_{\rm s}$ is about $4.4$ km for a $1.5\msun$ compact object.  We now find the optical depth for electron scattering to be
\be
\tau_{\rm es}\equiv\int^\infty_r \rho\kappa_{\rm es}dr=2R\dot{m}(rr_{\rm s})^{-1/2}, \label{eq:OptDep}
\ee
where $r$ is the distance from the compact object where the photon begins its journey.

From random walk, the time for the photon to diffuse over a distance $d$ is
\be
\tau_{\rm diff}=\frac{d}{c}\frac{d}{\lambda_{\rm es}}\simeq\frac{d}{c}\tau_{\rm es}, \label{eq:DiffTi}
\ee
where $\lambda_{\rm es}$ is the photon mean free path.  The approximation in the last step follows if we assume $d$ is large enough that most of the scatterings that the photon undergoes on its trip to infinity have already occurred by the time it has traveled distance $d$.  In order to obtain the photon trapping radius, we set this equal to the dynamical time, which is the time it would take for a piece of matter at position $r$ to travel the distance $d$ at its instantaneous speed at $r$, and approximates the time it takes for a photon to be advected inwards distance $d$:
\be
t_{\rm dyn}=\frac{d}{v(r)}=d\left(\frac{r}{2GM}\right)^{1/2}, \label{eq:DynTi}
\ee
where $v(r)$ is the free-fall velocity at $r$.  Using equation~\ref{eq:OptDep} for $\tau_{\rm es}$ and substituting $r_{\rm tr}$, the trapping radius, for $r$ we find
\be
r_{\rm tr}=2R\dot{m}. \label{eq:TrapR}
\ee
Since solution of the diffusion equation for radial diffusion in three dimensions decreases the diffusion time by a factor of $\pi^2/3$, we must likewise decrease $r_{\rm tr}$ by this factor.  We finally obtain
\be
r_{\rm tr}=0.6R\dot{m}=1.86\times10^9{\rm cm}. \label{eq:TrapR3d}
\ee
Any photon flux emitted much below $r_{\rm tr}$ is unable to diffuse upstream and thus can have no effect on the luminosity reaching the observer at infinity.

Chevalier and collaborators tool into account that neutrinos can carry away accretion energy and developed self consistent solutions for hypercritical accretion (Chevalier (1989), (1990); \cite{Hou91}).  In particular they find an expression for the radius of the accretion shock in terms of $\dot{M}$, for a neutron star.  We follow the derivation in Chevalier (1989, p.854) with small modification.  We first derive an expression for $p_{\rm ns}$, the pressure at the surface of the neutron star, in terms of $\dot{M}$ and $r_{\rm sh}$, the shock radius.  We then examine neutrino cooling near the surface of the neutron star, producing an equation in terms of $p_{\rm ns}$.  Insertion of our expression for $p_{\rm ns}$ gives $r_{\rm sh}$ in terms of $\dot{M}$.

Since the pressure is radiation dominated, the accretion envelope forms an $n=3$ ($\Gamma=4/3$) polytrope.  Thus, inside the shock
\be
\rho=\rho_{\rm sh}\left(\frac{r}{r_{\rm sh}}\right)^{-3};  p=p_{\rm sh}\left(\frac{r}{r_{\rm sh}}\right)^{-4};  v=v_{\rm sh}\left(\frac{r}{r_{\rm sh}}\right), \label{eqs:RhoPV}
\ee
where the subscript sh refers to the value at the shock front.  Because of the adiabatic compression by factor $(\Gamma+1)/(\Gamma-1)$,
\be
\rho_{\rm sh}=7\rho_0,\label{eq:Comp}
\ee
where $\rho_0$ is the density just outside the shock front.  We neglected the (small) decrease in $\Gamma$ because of increased ionization of the material going through the shock.  As in \citet{Blo86}, $\rho_0$ is calculated as follows:
\be
\rho_0=\frac{\dot{M}}{4\pi r_{\rm sh}^2 v_{in}}, \label{eq:Rho0}
\ee
where $v_{\rm in}$ is the free-fall velocity at the shock radius.  From conservation of mass flow across the shock front,
\be
v_{\rm sh}=-\frac{1}{7}v_{\rm in}.
\ee
Thus, the kinetic energy of the accreting matter is diminished by a factor of $49$; that is, it is almost entirely converted into thermal energy, so we can estimate the thermal energy density as
\be
\epsilon_{\rm sh}\simeq\frac{7}{2}\rho_0v_{\rm in}^2,
\ee
the factor 7 entering because of compression (equation~\ref{eq:Comp}.  In this derivation we neglected $v_{\rm sh}$ as compared with $v_{\rm in}$.  The small correction that would be generated had we taken account of the decrease in $\Gamma$ due to ionization across the shock in obtaining eq.~\ref{eq:Comp}.  Since the pressure is radiation dominated,
\be
p_{\rm sh}=\frac{1}{3}\epsilon_{\rm sh}\simeq\frac{7}{6}\rho_0v_{\rm in}^2.
\ee
Inserting equation~\ref{eq:Rho0} and $(2GM/r_{\rm sh})^{1/2}$ for $v_{\rm in}$, and using the second equation of~\ref{eqs:RhoPV}, we find the pressure at the surface of the neutron star
\be
p_{\rm ns}=1.86\times10^{-12}\dot{M}r_{\rm sh}^{3/2}{\rm dyn cm}^{-2}, \label{eq:Pns}
\ee
where $\dot{M}$ is expressed in g s$^{-1}$ and $r_{\rm sh}$ in cm.  We took the radius of the neutron star to be $\simeq10$ km.

The energy loss by neutrino pair production per unit volume is \citep{Dic72}
\be
\dot{\epsilon}_n=1.06\times10^{25}T^9C\left(\frac{\mu_e}{T}\right){\rm ergs cm}^{-3}{\rm s}^{-1}, \label{eq:dotEps}
\ee
where $\mu_e$ is the electron chemical potential, $T$ is in MeV, and $C(x)$ is a slowly varying function of $x$ which can be computed from the paper of \citet{Dic72}.  For $x=0$, $C=0.92$; we shall use this value, because the electrons are not very degenerate.  In the region where $\dot{\epsilon}_n$ is operative, $T\sim1$ MeV so that $e^+$, $e^-$ pairs as well as photons, contribute to the radiation pressure.  With $T$ in MeV, the photon blackbody energy density is
\be
w=1.37\times10^{26}T^4{\rm ergs cm}^{-3}.
\ee
Inclusion of the $e^+$, $e^-$ pairs multiplies this by a factor of $11/4$, and we divide by three to obtain the pressure
\be
p=1.26\times10^{26}T^4{\rm ergs cm}^{-3}. \label{eq:P}
\ee
Combining equations~\ref{eq:dotEps} and~\ref{eq:P} yields
\be
\dot{\epsilon}_{\rm n}=1.83\times10^{-34}p^{2.25},
\ee
where $\dot{\epsilon}_{\rm n}$ is in ergs cm$^{-3}$ s$^{-1}$ when $p$ is in ergs cm$^{-3}$.  The neutrino cooling is taken to occur within a pressure scale height $r_{\rm ns}/4$ of the neutron star, or in a volume of $\simeq\pi r_{\rm ns}^3$.  Energy conservation gives
\be
\pi r_{\rm ns}^3\times1.83\times10^{-34}p_{\rm ns}^{2.25}=\frac{GM\dot{M}}{r_{\rm ns}}, \label{eq:EnCons}
\ee
with, again, everything in cgs units.  Inserting equation~\ref{eq:Pns} into equation~\ref{eq:EnCons}, we solve for $r_{\rm sh}$:
\be
r_{\rm sh}\simeq6.4\times10^8\left(\frac{\dot{M}}{\msun{\rm yr}^{-1}}\right)^{-10/27}{\rm cm}.
\ee
The power $-10/27=-0.370$ is the same as that obtained by \citet{Hou91} using the accurate neutrino cooling function, and not that of Chevalier (1989).

The detailed calculation of \citet{Hou91} finds the only substantial correction to our schematic calculation to arise from general relativity, which can be taken into account by multiplying the expression for $r_{\rm sh}$ by $0.4$.  Thus,
\be
r_{\rm sh}\simeq2.6\times10^8\left(\frac{\dot{M}}{\msun{\rm yr}^{-1}}\right)^{-0.370}{\rm cm}=6.3\times10^9{\rm cm},
\ee
where we have used equation~\ref{eq:dotM} for $\dot{M}$.

\citet{Bro94} go on to calculate the critical time after which the neutron star left behind by SN 1987a should be visible and find it is less than a year after the explosion.  Up to this moment, no neutron star has been observed.  Nevertheless the hypercritical accretion onto a black hole will be just as efficient, if not more than onto a neutron star, since the black hole has no surface against which matter would hit and produce radiation pressure, but only an event horizon which it would ram through without resistance.


\end{document}